\documentclass[twocolumn,prb,preprintnumbers,amsmath,amssymb,sectsty]{revtex4}
\usepackage{amsmath,amssymb,color,graphicx,epsfig}
\usepackage{rotating}
\usepackage{subfigure}
\usepackage{dcolumn}
\usepackage{bm}

\newcommand{\abs}[1]{|#1\rvert}
\newcommand{\ds}{\,\text{d}s}

\newcommand{\dt}{\,\text{d}t}

\newcommand{\dr}{\,\text{d}r}

\newcommand{\dN}{\,\text{d}N}
\newcommand{\dtheta}{\,\text{d}\theta}
\newcommand{\deta}{\,\text{d}\eta}
\newcommand{\etal}{\it{\text{et al.}}}
\def\bb{{\bf b}}\def\bd{{\bf d}}
\def\be{{\bf e}}

\def\bn{{\bf n}}
\def\br{{\bf r}}\def\bt{{\bf t}}
\def\bx{{\bf x}}

\def\beqn{\begin{equation}}
\def\eeqn{\end{equation}}

\begin{document}

\title{Elastic free-energy of wormlike micellar chains: theory and suggested experiments}

\author{{Meisam Asgari}}

\affiliation{Department of Mechanical Engineering, McGill University,\\ 817 Sherbrooke Street West, Montr\'eal, QC H3A0C3, Canada\\Fax: 514 398 7365; E-mail: meisam.asgari@mail.mcgill.ca}

\begin{abstract}

\noindent {The extensive application of surfactants motivates comprehensive and predictive theoretical studies that improve our understanding of the behaviour of these complex systems. In this study, an expression for the elastic free-energy density of a wormlike micellar chain is derived taking into account interactions between its constituent molecules. The resulting expression incorporates the sum of a quadratic term in the curvature and a quadratic term in the torsion of the centerline of wormlike micelle and thus resembles free-energy density functions for polymer chains and DNA available in the literature. The derived model is applied on a wormlike micelle in the shape of a circular arc, open or closed. A detailed application of the derived model on wormlike micelles of toroidal shape, along with employing necessary statistical-thermodynamical concepts of self-assembly, is performed, and the results are found to be consistent with the ones available in the literature. Steps towards obtaining the material parameters through experiments are suggested and discussed.}

\end{abstract}

\maketitle

\section{Introduction}

A surfactant molecule consists of two main parts: a hydrophobic tail and a polar hydrophilic head-group. When present in solution at sufficiently high concentrations, surfactant molecules self-assemble into various supramolecular structures that shield the hydrophobic tails from contact with the ambient solution (Israelachvili~\cite{israelachvili2011intermolecular,israelachvili1976theory}). These structures include spherical micelles, short cylindrical micelles, long cylindrical micelles called wormlike micelles, bilayers, and closed bilayers or vesicles (Cates and Candau~\cite{cates1990statics} and Discher {\etal}~\cite{discher1999polymersomes}). 
A collection of spherical micelles may undergo uniaxial growth and coalesce into short rod-like cylindrical micelles (Cui {\etal}~\cite{cui2007block} and Groswasser {\etal}~\cite{bernheim2000sphere}). The ends of a cylindrical micelle are capped by hemispheres (Porte {\etal}~\cite{porte1986morphological}). Adding more molecules at certain temperatures and osmotic pressures leads to the formation of wormlike micelles (May and Ben-Shaul~\cite{zana2007giant} and Dam {\etal}~\cite{van2004direct}). The energy required to create two end-caps from a very long cylindrical micelle is called the scission energy (Oelschlaeger {\etal}~\cite{oelschlaeger2008linear}). If this energy is sufficiently large and the volume fraction of the surfactant molecules is sufficiently low, the semiflexible micelles may fuse to minimize the number of end caps (Cates and Candau~\cite{cates1990statics}). 

In recent years, considerable attention have been paid to investigating the rheological properties of surfactant-based micellar solutions (Boek {\etal},~\cite{boek2007flow} Lerouge and Berret,~\cite{lerouge2010shear} and Spenley {\etal}~\cite{spenley1993nonlinear}). Some examples include soap solutions including micellar globules and adhesives formulated with block copolymers (Won {\etal}~\cite{won1999giant}). Due to their novel properties, such solutions have found applications as heat-transfer fluids, hard-surface cleaners, liquid dish-washing detergents, drag-reducing agents in pipelines, and fracking fluids (C\'ecile,~\cite{dreiss2007wormlike} Yang,~\cite{yang2002viscoelastic} and Komura and Safran~\cite{komura2001scaling}).

Wormlike micelles may be regarded as mesopolymers since they impart elasticity (Padding and Boek~\cite{padding2004evidence} and Berret~\cite{berret2006rheology}) and are subject to breaking and re-shaping (Boek {\etal},~\cite{boek2005mechanical} Zhou {\etal},~\cite{zhou2014wormlike} and Germann {\etal}~\cite{germann2013nonequilibrium,germann2014investigation}). They may also become entangled or form branched structures (Andreev and Victorov~\cite{andreev2007simple}). As a result, a solution in which they are suspended may have viscoelastic behaviour (Acharya and Kunieda,~\cite{acharya2003formation,acharya2006wormlike} Raghavan {\etal}~\cite{raghavan2002wormlike} and Yang~\cite{yang2002viscoelastic}). For this reason, many rheologists focus on understanding the effective non-Newtonian behavior of micellar solutions (Kuperkar {\etal}~\cite{kuperkar2008viscoelastic} and Radulescu {\etal}~\cite{radulescu2003time}). That behavior is generally influenced by micro-structural changes that occur during flow (Khatory {\etal}~\cite{khatory1993linear}). The present paper focuses on the energetics of an individual closed or open wormlike micelle, leaving aside questions related to dissipative interactions in flowing solutions.

Wormlike micelles are often characterized by their persistence length. This quantity can be obtained from the bending free-energy (Nettesheim and Wagner,~\cite{nettesheim2007fast} and Schubert {\etal}~\cite{schubert2003microstructure}). The free-energy density therefore has a significant role in studying these materials. May {\etal}~\cite{may1997molecular} considered the extent to which the bending elasticity of wormlike micelles influences their tendency to join and form branched structures. In their work, the free-energy density of a wormlike micelle is comprised of the chain conformational free-energy, the end-cap energies, and the hydrocarbon-water interfacial energy. Their findings indicate that the energy change associated with the formation of a junction between one micellar end-cap and the cylindrical body of another micelle is small relative to the free energy of an end cap.

The first proposed conformational free-energy density function for a wormlike micelle was motived by the Canham--Helfrich~\cite{canham1970minimum,helfrich1973elastic} elastic free-energy density for a lipid vesicle, 
\begin{equation}
\label{Helf. En.}
\psi=\gamma_\circ+\frac{1}{2}k_c(H-H_{\circ})^2+\bar{k}_cK,
\end{equation}
where $\gamma_\circ$ is the surface tension, $H$ and $K$ are the mean and Gaussian curvatures of the surface, $k_c$ and $\bar{k}_c$ are the splay and saddle-splay moduli, and $H_{\circ}$ is the spontaneous mean curvature, namely the mean curvature of the natural, local shape of the bilayer that has been attributed to the difference between the volumes of the head and tails (Tang and Carter~\cite{tang2013branching}).

Lauw {\etal}~\cite{lauw2003self} used self-consistent field-theory to estimate the free-energy of a closed toroidal wormlike micelle comprised of nonionic surfactant molecules. This relied on specializing the Helfrich expression~\eqref{Helf. En.} to a toroidal geometry. 
Proceeding similarly, Bergstr\"om~\cite{bergstrom2007bending, bergstrom2008thermodynamics} investigated the effect of the splay modulus on the size and shape of toroidal micelles and their stability.

An alternative approach to express the free energy of a wormlike micelle is based on the well-established molecular-statistical perspective for modeling self-assembled aggregates. In such an approach, the free energy per amphiphile is considered as the sum of various contributions including hydrophobic effect, chain conformational entropy, electrostatic effects (obtained from linearized Poisson--Boltzmann equation), and head-group repulsion (Puvvada and Blankschtein,~\cite{puvvada1990molecular} Nagarajan and Ruckenstein,~\cite{nagarajan1991theory} and May {\etal}~\cite{may1997molecular}). According to Bergstr\"om,~\cite{bergstrom2006model} the aforementioned contributions can be accurately described by applying Helfrich's curvature elasticity theory~\eqref{Helf. En.}.

Wormlike micelles have diameters on the order of 3--5 nanometers. The contour length of a wormlike micelle need not be fixed and may be sensitive to environmental conditions. The length of a wormlike micelle may be extremely large relative to its cross-sectional diameter, reaching 3600 nanometers (C\'ecile~\cite{dreiss2007wormlike}) or as much as one millimeter (Padding~\cite{padding2004evidence,padding2004influence}). The ratio of the length of a polymer chain to its thickness can be up to 500 or more (Roiter and Minko~\cite{roiter2005afm}). The same ratio for a typical human hair ranges between $10^2$ and $10^4$. The length to diameter ratio of a wormlike micelle exceeds the aforementioned ratios for polymer chains or human hair. Considering the geometric analogy between wormlike micelles and polymer chains, and the application of rod-like models for polymer chains (Kratky and Porod,~\cite{kratky1949rontgenuntersuchung} and Bugl and Fujita~\cite{bugl2003dynamics}), an approach closer to those used to model polymers seems to be worthy of consideration.

As previously mentioned, the notion of free-energy density is central to the study of wormlike micelles. For instance, with an expression for the free-energy density of such a micelle, equilibrium configurations can be explored. Here, a free-energy density function is derived for wormlike micelles. To do so, an approach based on accounting for the interactions between the constituent molecules of the micelle is adopted. This approach is largely based on ideas developed by Keller and Merchant,~\cite{keller1991flexural} who showed that the surface energy of a substance generally includes a bending term and, thus, that the surface may exhibit elastic resistance to bending even in the absence of stretching. This contribution to the free energy is obtained in terms of the molecular density and interaction potential. In a recent application of the work of Keller and Merchant,~\cite{keller1991flexural} Seguin and Fried~\cite{seguin2014microphysical} derived the Canham--Helfrich free-energy density for a lipid vesicle. In so doing, they modeled the lipid molecules comprising the vesicle by one-dimensional rigid rods. For simplicity, they neglected molecular tilt relative to the orientation of the bilayer and interactions between the bilayer and the solution.

To determine the free-energy density of a wormlike micelle at a position $\bx$, we account for the interactions between all surfactant molecules within a cutoff distance $d$ from the molecules at $\bx$. For simplicity, we model surfactant molecules as one-dimensional rigid rods and assume that these rods are  perpendicular to the centerline of the wormlike micelle. Additionally, interactions between the surfactant molecules comprising the wormlike micelle and the surrounding solution are neglected. Our derivation relies on a Taylor series expansion with respect to a dimensionless parameter $\delta:=d/\ell\ll1$, where $\ell$ is the smallest radius of curvature that the centerline of a wormlike micelle is capable of exhibiting. The free-energy of an open wormlike micelle results on integrating this density over the centerline of the wormlike micelle and adding the end-cap energies.

This article is organized as follows: In Section~\ref{sectmain2}, modeling assumptions and geometrical considerations are briefly presented. The section is followed by introducing the distribution function of the molecules along the body of the wormlike micelle. Section~\ref{sectmain3} includes the calculation of the free-energy density along the tubular body of the wormlike micelle, followed by obtaining the free-energy of closed and open wormlike micelles. Section~\ref{sectmain4} is concerned with the application of the model to a wormlike micelle in the shape of a circular arc, open or closed. A detailed application of the model on wormlike micelles of toroidal geometry along with statistical-thermodynamical concepts and suggested experimental procedure towards obtaining the material parameters are provided in Section~\ref{sectmain6}. Finally, the key findings and implications of this study are summarized and discussed in Section~\ref{sectmain7}. Details of the various derivations are presented in the Appendix.
\section {Modeling assumptions}
\label{sectmain2}
According to the assumptions below, the body of a wormlike micelle is envisioned as a tubular domain with centerline ${\cal C}$, of length $L$, and constant cross-sectional radius $a$:
\begin{enumerate}
{\item The surfactant molecules that comprise the micelle have identical physiochemical properties.\label{as1}}
{\item Each surfactant molecule can be modeled as a one-dimensional rigid rod of length $a$.\label{as2}}
{\item The surfactant molecules at any point of ${\cal C}$ are arranged perpendicular to ${\cal C}$ with a uniform angular distribution.\label{as3}}
{\item The length $L$ of the wormlike micelle far exceeds its radius $a$, namely the length $a$ of a single molecule.\label{as4}}
{\item The minimum radius of curvature that a wormlike micelle can support at each point is denoted by $\ell$.\label{as5}}
\end{enumerate}

As a consequence of assumptions~\ref{as2} and~\ref{as3}, the head-groups of surfactant molecules lie on a tubular surface of constant circular cross-section. Assumption~\ref{as3} is based on the observation that the hydrophobic tails of surfactant molecules are oriented (on average) along the normal of the tubular interface (May {\etal}~\cite{may1997molecular} and Shikata {\etal}~\cite{shikata1994rheo}). Assumption~\ref{as4} allows the wormlike micelle to be identified with its centerline ${\cal C}$. Assumption~\ref{as5} refers to the fact that the tubular body of the micelle may adopt various shapes; however, being made up of molecules of a finite size, it cannot support arbitrarily large curvatures. Moreover, the tubular body does not overlap (or even contact) itself. On this basis, there is an upper bound for the curvature (or, equivalently, a lower bound for the radius of curvature) which ensures that contact/overlap does not occur. We denote such radius of curvature by $\ell$.

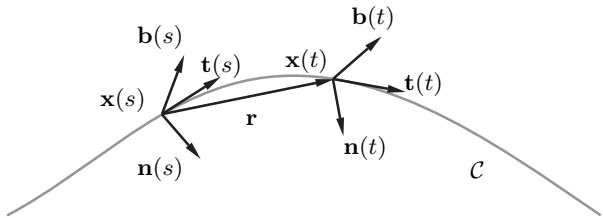
\begin{figure} [b]
 \centering
 \includegraphics [height=0.95in] {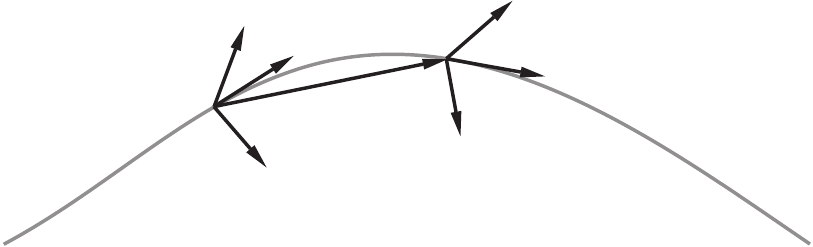}
   \put(-120,57){\small$\bx(t)$}   
   \put(-135,35){\br}
   \put(-176,16){\small$\bn(s)$}
   \put(-176,66){\small$\bb(s)$}   
   \put(-190,42){\small$\bx(s)$}
   \put(-50,15){\small${\cal C}$}
   \put(-152,55){\small$\bt(s)$}   
   \put(-95,73){\small$\bb(t)$}
   \put(-75,48){\small$\bt(t)$}   
   \put(-98,24){\small$\bn(t)$}
   \caption{\footnotesize Two positions $\bx(s)$ and $\bx(t)$ on the centerline of a wormlike micelle with the intermolecular vector $\br$, and the Frenet frame at those positions.}
    \label{f4}
\end{figure}
\subsection{Preliminary geometrical considerations}
Let ${\cal C}$ denote the centerline of the tubular domain that represents the wormlike micelle and consider an arclength parametrization
\begin{equation}
\label{curve}
{\cal C}=\{\bx:\bx=\bx(s),0\le s\le L\}.
\end{equation}
Using a prime to denote differentiation with respect to the arclength $s$, it follows that
\begin{equation}
\label{constraint}
\abs{\bx'}=\abs{\frac{d\bx}{\ds}}=\Big({\frac{d\bx}{\ds}\cdot\frac{d\bx}{\ds}}\Big)^{1/2}=1
\end{equation}
and, also, that
\begin{equation}
\label{constraint2}
\bx'\cdot\bx''=0,\quad\text{and}\quad\abs{\bx'\times\bx''}=\abs{\bx''}. 
\end{equation}
The unit tangent $\bt$, unit normal $\bn$, and unit binormal $\bb$ of the Frenet frame $\{\bt,\bn,\bb\}$ of ${\cal C}$ are given in terms of the arclength parametrization $\bx$ of ${\cal C}$ by (Figure~\ref{f4})
\begin{equation}
\begin{split}
&\bt=\bx',\quad
\bn=\frac{\bx''}{\abs{\bx''}},
\quad\text{and}\quad
\bb=\frac{\bx'\times\bx''}
{\abs{\bx''}}.
\end{split}
\label{ex11}
\end{equation} 
Moreover, the curvature $\kappa$ and torsion $\tau$ of ${\cal C}$ are given by 
\begin{equation}
\kappa=\abs{\bt'}=\abs{\bx''},
\quad\text{and}\quad
\tau=\pm\abs{\bb'}=\frac{\bx'\cdot(\bx''\times\bx''')}{\abs{\bx''}^2}.
\label{curvtor}
\end{equation}

Let $s$ belong to the interval $(0,L)$, so that $\bx(s)$ is interior to ${\cal C}$. Consider a molecule at the position $\bx(s)$ with orientation $\theta$ measured counterclockwise from the line determined by the unit normal $\bn(s)$. The director of this molecule is given by the unit vector $\bd(s,\theta)$. According to assumption~\ref{as3}, the director $\bd(s,\theta)$ of the molecule at ${\bf x}(s)$, can be expressed as a linear combination 
\begin{equation}
\label{ex0}\bd(s,\theta)=(\cos{\theta})\,\bn(s)+(\sin{\theta})\,\bb(s),
\end{equation} 
of unit normal $\bn(s)$ and unit binormal $\bb(s)$ (Figure~\ref{f3}). Similarly, the director $\be(t,\eta)$ of the molecule at position $\bx(t)$ can be expressed as a linear combination of $\bn(t)$ and $\bb(t)$ by (Figure~\ref{f3})
\begin{equation}
\label{ex000}\be(t,\eta)=(\cos{\eta})\,\bn(t)+(\sin{\eta})\,\bb(t).
\end{equation} 

\subsection{Molecular distribution function}
The way that surfactant molecules are distributed along the tubular body of the wormlike micelle, has an effect in describing how such molecules interact. The molecular distribution function describes how such molecules are distributed along ${\cal C}$.

The distribution of the molecules at a generic point $s$ in the open interval $(0,L)$ along ${\cal C}$ is denoted by $f>0$. It follows that the integral
\begin{equation}
\label{eqn11}
\int_{0}^{L}\mskip-8mu\int_{0}^{2\pi}f(s)\,\dtheta\,\ds
\end{equation}
represents the total number of surfactant molecules comprising the tubular body of the wormlike micelle. In view of assumption~\ref{as3} in Section~\ref{sectmain2}, the angular distribution of surfactant molecules alone ${\cal C}$ is uniform. Consequently, \eqref{eqn11} yields
\begin{equation}
\label{eqn122}
2\pi\int_{0}^{L}f(s)\ds.
\end{equation}

\section{Free-energy of a wormlike micelle}\label{sectmain3}

Wormlike micelles are either closed or open. An open wormlike micelle possesses two end caps. Branched wormlike micelles, however, possess more than two end caps along with junctions which may contribute additional free energy (Jain and Bates~\cite{jain2003origins} and Dam {\etal}~\cite{van2004direct}). Here, we first derive the free-energy density on the body of the wormlike micelle. The net free-energy of a closed wormlike micelle is simply found by integrating the derived free-energy density over the body of the micelle. For an open wormlike micelle, the free energy corresponding to the end caps should be added to the net free-energy of the body. For simplicity, we restrict attention to wormlike micelles that do not possess branches and, thus, possess only two end caps. 
\begin{figure} [t]
 \centering
 \includegraphics [height=2.2in] {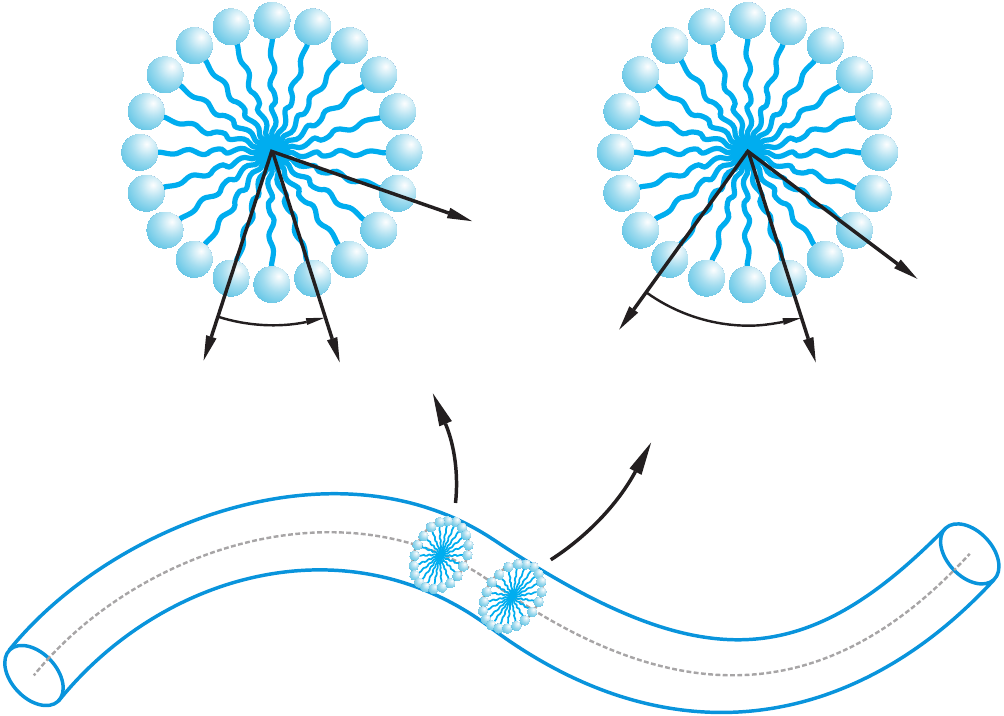}
   \put(-125,5){$\bx(t)$} 
   \put(-144,17){$\bx(s)$}
   \put(-52,71){$\be(t,\eta)$}
   \put(-160,71){$\bd(s,\theta)$}
   \put(-116,105){$\bb(s)$}
   \put(-190,70){$\bn(s)$}   
   \put(-50,11){${\cal C}$}
   \put(-167,79){$\theta$}   
   \put(-93,77){$\bn(t)$}
   \put(-67,80){$\eta$}   
   \put(-19,95){$\bb(t)$}
 \caption{\footnotesize Geometry of a part of the tubular body of a wormlike micelle based on simplifying assumptions. The head groups of the molecules at $\bx(s)$ must lie on the boundary of a disk (of radius $a$) in the plane spanned by $\bn(s)$ and $\bb(s)$.}
  \label{f3}
\end{figure}
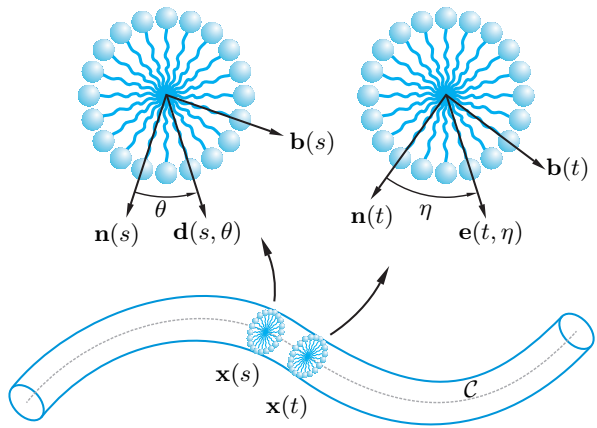

\subsection{Free-energy density of the tubular body of the wormlike micelle}

The interaction energy between two rod-like structures is based on three ideas, the first of which entails considering the molecules as one-dimensional rigid rods, the second embodies the notion that the rods are made up of material points that interact with each other, and the third being the principle of material frame-indifference, which states that the constitutive relations describing the internal interactions between the parts of a system should not depend on the external frame of reference used to describe them. This principle places restrictions on the constitutive equations (Truesdell and Noll~\cite{truesdell2004non}).

Assume that the interaction between two molecules is governed by an energy function depending on the location of the two molecules and the directors indicating the orientation of the molecules.~\cite{seguin2014microphysical,de2014modeling} Consider two molecules at positions $\bx(s)$ and $\bx(t)$ on ${\cal C}$, as depicted schematically in Figure~\ref{f3}. From \eqref{ex0} and~\eqref{ex000}, their directors are of the form $\bd(s,\theta)$ and ${\be}(t,\eta)$, respectively. Let 
\begin{equation}
\label{2121}\Omega\big(\bx(s),\bx(t),\bd(s,\theta),\be(t,\eta)\big),
\end{equation}
denote the interaction energy between these two molecules. An interaction energy of the form~\eqref{2121} may encompass different effects such as steric or electrostatic interactions between surfactant molecules. Following Keller and Merchant,~\cite{keller1991flexural} assume that the interaction energy between two molecules separated by a distance greater than some fixed cutoff distance $d$ vanishes. To express this differently, assume that only molecules within a distance $d$ may interact. On denoting the relative position vector between a pair of molecules at $\bx(s)$ and $\bx(t)$ by $\br=\bx(s)-\bx(t)$, it follows that $\Omega$ satisfies
\begin{equation}
\label{condition}\abs{\br}>d \Longrightarrow \Omega\big(\bx(s),\bx(t),\bd(s,\theta),\be(t,\eta)\big)=0. 
\end{equation}
It is assumed that the cutoff distance $d$ is small relative to the minimum radius of curvature $\ell$ introduced in assumption~\ref{as5}, so that $d\ll{\ell}$ or,  
\begin{equation}
\delta:=\frac{d}{\ell}\ll1.
\end{equation}
Granted that $\Omega$ is frame indifferent, it must obey
\begin{align}
\label{212}\notag&\Omega\big(\bx(s),\bx(t),\bd(s,\theta),\be(t,\eta)\big)\\
&\qquad=2\,\hat\Omega\big(\frac{\abs{\br}^2}{\delta^2},\br\cdot\bd(s,\theta),\br\cdot\be(t,\eta),\bd(s,\eta)\cdot\be(t,\eta)\big),
\end{align}
where a factor of 2 has been introduced to simplify subsequent calculations. The quantity $\hat\Omega$ on the right-hand side of~\eqref{212} is a function of four scalar arguments. These arguments include the length of the intermolecular vector $\br$, and the dot products $\br\cdot\bd$ and $\br\cdot\be$ between the directors and that vector and the dot product $\bd\cdot\be$ between the directors. Such dot products are related to the angles $\alpha_1$ and $\alpha_2$ between the relative position vector $\br$ and the directors $\bd$ and $\be$, and also to the angle $\alpha_3$ between the two directors $\bd$ and $\be$, by
\begin{equation}
{\hat\br}\cdot\bd=\cos\alpha_1,
\quad
{\hat\br}\cdot\be=\cos\alpha_2,
\quad
\bd\cdot\be=\cos\alpha_3.
\end{equation}  
with $\hat\br$ being the unit vector corresponding to the intermolecular vector $\br$. According to~\eqref{condition}, $\Omega$ depends implicitly upon $d$ through $\delta$, while $\hat\Omega$ does not depend upon $d$. The term $\delta^2$ in the denominator of the first argument on the right-hand side of \eqref{212} is introduced to remove that dependence.
It follows from \eqref{condition}--\eqref{212} that $\hat\Omega$ satisfies
\begin{equation}
\label{ieradius}
{s>\ell\Longrightarrow\hat\Omega(s^2,s\delta\cos\alpha_1,s\delta\cos\alpha_2,\cos\alpha_3)=0.}
\end{equation}

The total free-energy $E_{\text{tot}}$ of the tubular body of the wormlike micelle is given by
\begin{align}
\label{energy}
\notag E_{\text{tot}}&=\int_{0}^{L}\mskip-8mu\int_{0}^{L}\mskip-8mu\int_{0}^{2\pi}\mskip-12mu\int_{0}^{2\pi}\frac{1}{2}\Omega\big(\bx(s),\bx(t),\bd(s,\theta),\be(t,\eta)\big)\\
&\qquad\qquad\qquad\qquad\qquad{f(s)}{f(t)}\,\dtheta\deta\dt\ds.
\end{align}
To reiterate, $\Omega\big(\bx(s),\bx(t),\bd(s,\theta),\be(t,\eta)\big)$ is the interaction energy between a molecule at position $\bx(s)$ with orientation $\bd(s,\theta)$ and a molecule at $\bx(t)$ with orientation $\be(t,\eta)$. Whereas $f(s)$ gives the density of the molecules at $\bx(s)$, $f(t)$ gives the same quantity at $\bx(t)$. By integrating from $0$ to $2\pi$ with respect to the variables $\theta$ and $\eta$, the interaction energy between all of the molecules of the two cross-sections at $\bx(s)$ and $\bx(t)$ is taken into account. The integrand in~\eqref{energy} is multiplied by $1/2$ because otherwise it double counts the interaction energy between each pair of molecules. By integrating over $t$ from $0$ to $L$, the interaction of all the molecules of the other cross-sections with the molecules located in the cross-section at the position $\bx(s)$ is incorporated. The second integration, over $s$ from $0$ to $L$, ensures that the free energy of each cross-section is counted.

The net free-energy function $E_{\text{tot}}$ of the tubular body of the wormlike micelle is related to the free-energy density $\psi$ by
\begin{equation}
E_{\text{tot}}=\int_{0}^{L}\psi\,\ds.
\end{equation}
\begin{figure} [!b]
 \centering
 \includegraphics [height=1in] {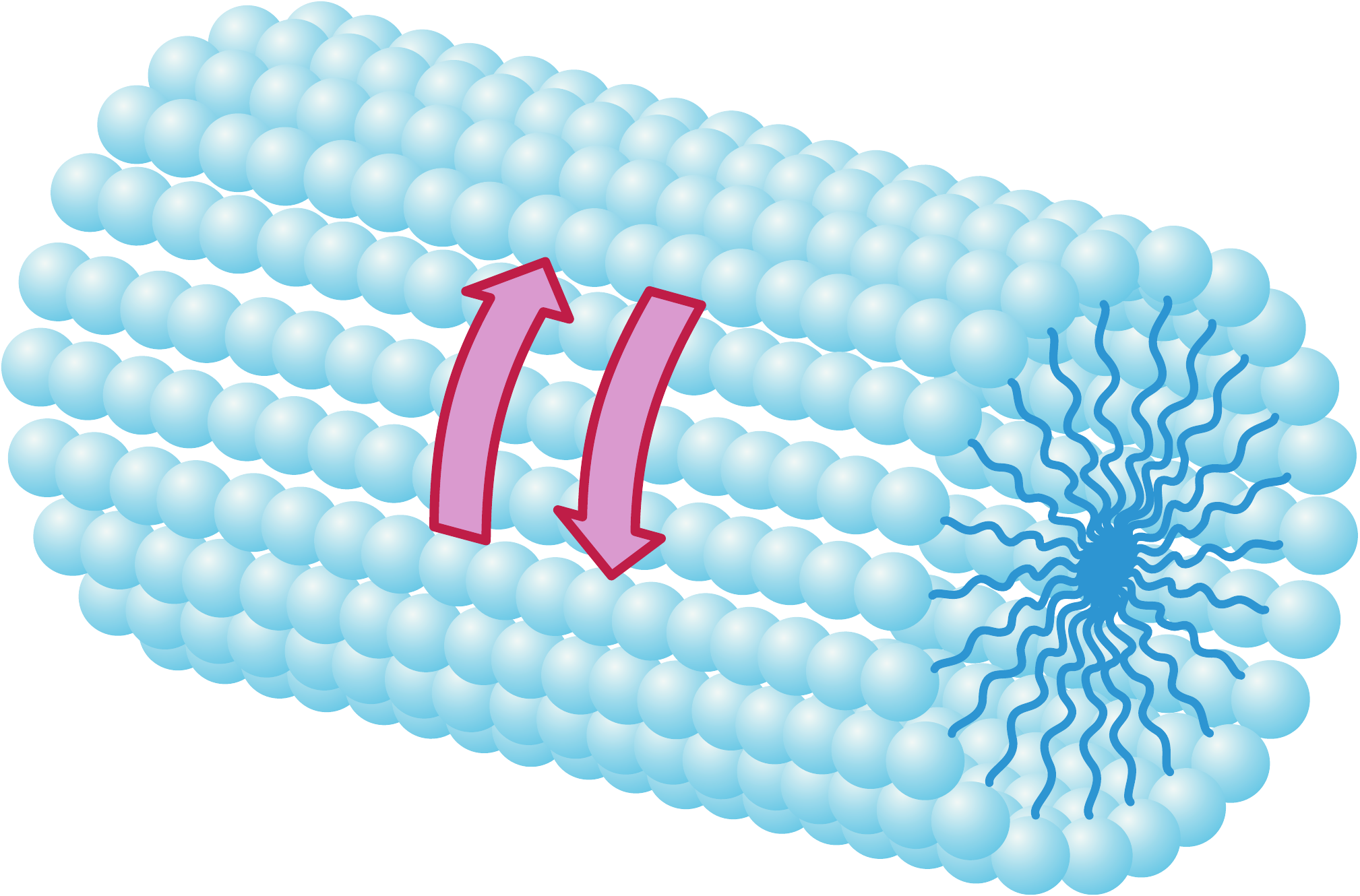}
  \caption{\footnotesize Schematic of a small section of a wormlike micelle. Since the molecules can rotate freely, no energetic cost in incurred by twisting.}
\label{f6}
\end{figure}

By~\eqref{212}--\eqref{energy}, the free-energy density $\psi$ of the wormlike micelle at a position $\bx=\bx(t_{\circ})$ on ${\cal C}$ is, up to an arbitrary additive constant, given by
\begin{align}
\label{e11}
\notag \psi&=\int_{0}^{L}\mskip-8mu\int_{0}^{2\pi}\mskip-11mu\int_{0}^{2\pi}\hat\Omega\Big(\frac{\abs{\bar\br(t)}^2}{\delta^2},{\bar\br(t)}\cdot\bd(t_{\circ},\theta),{\bar\br(t)}\cdot\be(t,\eta),\\
&\qquad\qquad\qquad\bd(t_{\circ},\theta)\cdot\be(t,\eta)\Big)f(t_{\circ}) f(t)\,\dtheta \deta \dt,
\end{align}
where $\bar\br(t)=\bx(t_{\circ})-\bx(t)$ is the intermolecular vector.

Upon performing the Taylor expansion of the right-hand side of \eqref{e11} up to two derivatives, the specific steps of which appear in Appendix, the final form of the free-energy density for a wormlike micelle is found to be 
\begin{equation}
\label{finalform2}
\psi=\psi_{\circ}+k_1\kappa^2+k_2\tau^2,
\end{equation}
which includes the sum of a quadratic term in curvature $\kappa$ and a quadratic term in torsion $\tau$ of the centerline ${\cal C}$ of the micelle, defined in \eqref{curvtor}. Notice that the free-energy density $\psi$ in~\eqref{finalform2} is independent of the size and shape of the wormlike micelle. Thus, according to our theory, the free energy of a wormlike micelle in an equilibrated system is completely determined by the parameters $\psi_\circ$, $k_1$ and $k_2$.

The term $\psi_{\circ}$ in~\eqref{finalform2} is insensitive to the shape of the wormlike micelle. Since the molecular distribution function $f$ has implicit dependence upon effects like temperature, concentration, and electromagnetic fields, these effects may be encompassed in $\psi_{\circ}$ and in the moduli $k_1$ and $k_2$. The role of the term $\psi_\circ$ in~\eqref{finalform2} is similar to that of $\gamma_\circ$ in Helfrich's free-energy density~\eqref{Helf. En.}. However, $\gamma_\circ$ in~\eqref{Helf. En.} is not an independent parameter, and is determined by the condition of constant overall surfactant concentration (Bergstr\"om~\cite{bergstršm2009bending}).

The flexural (or bending) rigidity $k_1$ in~\eqref{finalform2} represents the resistance of the wormlike micelle against deviations from a uniform curvature. It further describes the stability and stiffness of the wormlike micelle (Gradzielski~\cite{gradzielski2003kinetics}). Similarly, the torsional rigidity $k_2$ denotes the resistance of the wormlike micelle against moving out of the osculating plane of the centerline ${\cal C}$ at each point. Notice that the moduli $k_1$ and $k_2$ depend upon the size and composition of surfactant molecules comprising the micelle, the distribution of the molecules, temperature, and the concentration of the solution (Bergstr\"om~\cite{bergstrom2006model}).

\begin{figure} [!t]
 \centering
 \includegraphics [height=1.1in] {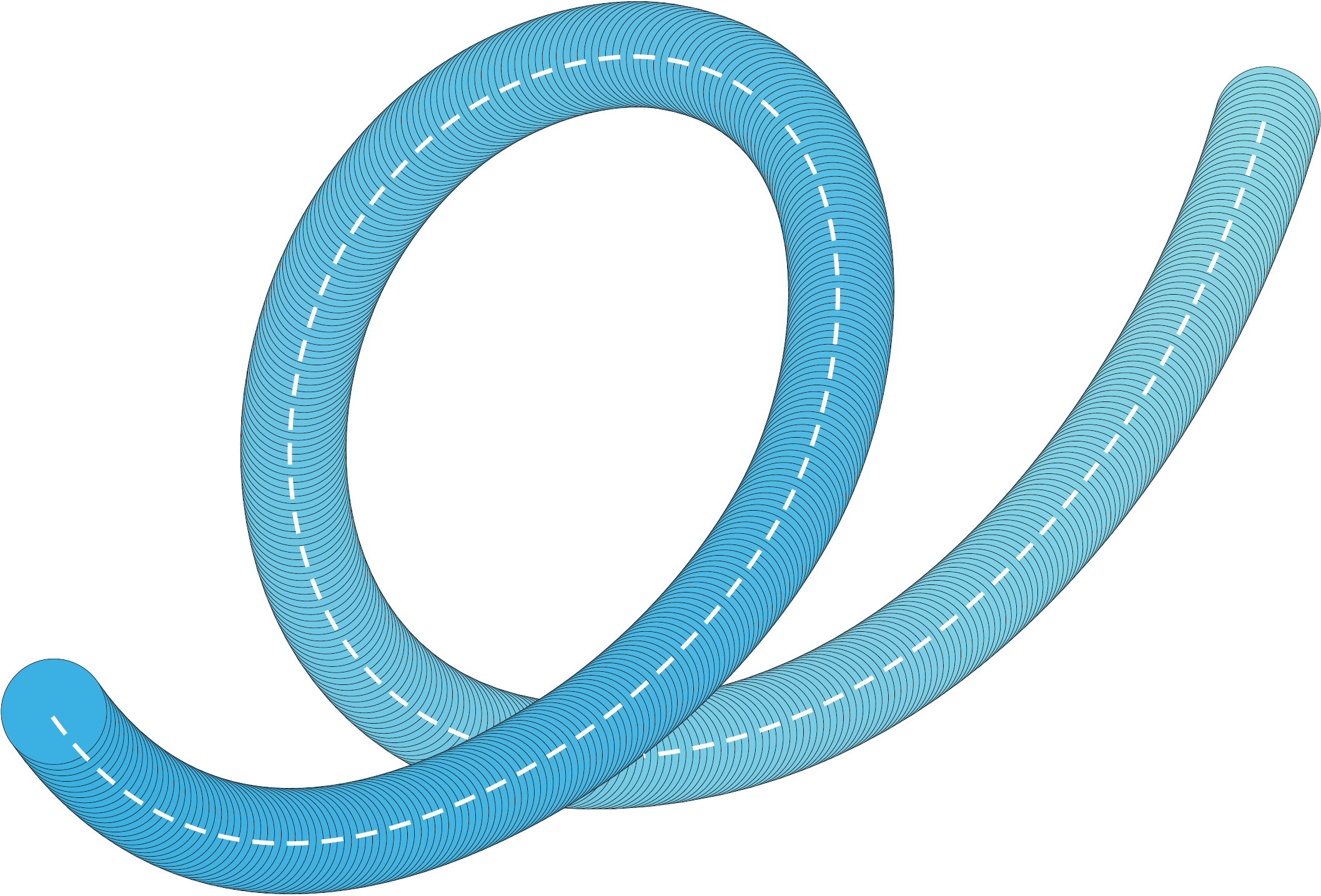}
  \caption{\footnotesize Schematic of a wormlike micelle that possesses torsion.}
\label{figure4}
\end{figure}

The torsion $\tau$ of ${\cal C}$ should not be confused with the notion of the twist between different cross-sections of ${\cal C}$. Since the molecules comprising the wormlike micelle may adjust their positions in response to the twist between adjacent cross-sections, as shown in Figure~\ref{f6}, the expression~\eqref{e11} for the free-energy density $\psi$ contains no contribution related to twist. In contrast, $\tau$ measures the turnaround of the binormal $\bb$ of ${\cal C}$ or, equivalently, the tendency of ${\cal C}$ to move out of a given osculating plane (Figure~\ref{figure4}). Thus, $\tau$ vanishes for plane curves and is constant for helical curves. As defined in~\eqref{curvtor}$_2$, $\tau$ is positive (negative) for a right-handed (left-handed) helix.

The expression in~\eqref{finalform2}, which is known in the context of the elasticity of bent rods (Landau and Lifshitz~\cite{landau1975elasticity}), resembles free-energy density functions for polymer chains (Kratky and Porod,~\cite{kratky1949rontgenuntersuchung} Bugl and Fujita,~\cite{bugl2003dynamics} and Liu {\etal}~\cite{liu2011statistical}) and those for DNA (Marko and Siggia,~\cite{marko1994bending,marko1995stretching} and Balaeff {\etal}~\cite{balaeff2006modeling}). However, in contrast to those models, \eqref{finalform2} contains evidence of neither intrinsic curvature nor intrinsic torsion. For a wormlike micelle consisting of more than one type of surfactant molecule, the presence of molecules with different head-group or tail conformations might lead to intrinsic curvature or intrinsic torsion. In the present setting, such effects are ruled out by our first modeling assumption.

\subsection{Free energy of a closed wormlike micelle}
For a closed wormlike micelle, the centerline ${\cal C}$ is closed. Integrating the free-energy density function in \eqref{finalform2} 
over ${\cal C}$ yields the total free-energy,
\begin{equation}
\label{total-energy}
E_{\text{tot}}=\int_{{\cal C}}\big(\psi_{\circ}+k_1\kappa^2+k_2\tau^2\big)\ds,
\end{equation}
of a closed wormlike micelle with centerline ${\cal C}$.
\subsection{Free energy of an open wormlike micelle}
Consider now an open wormlike micelle with two end caps. The net free-energy of such a wormlike micelle is obtained by adding the free energy of the tubular body and that of the two end caps. The former is obtained by integrating the free-energy density~\eqref{finalform2} over ${\cal C}$. The latter is obtained by 2 times the free energy $E_{\text{cap}}$ corresponding to an end cap. The result is given by the sum 
\begin{equation}
\label{open WM}
E_{\text{tot}}=\int_{{\cal C}}\big(\psi_\circ+k_1\kappa^2+k_2\tau^2\big)\ds+2E_{\text{cap}}.
\end{equation}

\section{Illustrative examples}\label{sectmain4}
In this section, we investigate the application of our model on two examples of wormlike micelles, the first of which an open wormlike micelle in the shape of a circular arc, and the second, a closed wormlike micelle in the shape of a torus. Bearing in mind the relatively high difference between the free energy associated with the end caps and that associated with the tubular body of the micelle, we then explore the condition under which the open wormlike micelle in the shape of the circular arc tends to close itself to form a toroidal micelle.

\subsection{Example 1: open wormlike micelle}{\label{sectmain4.1}}

\begin{figure}[b]
 \centering
 \includegraphics [height=.78in] {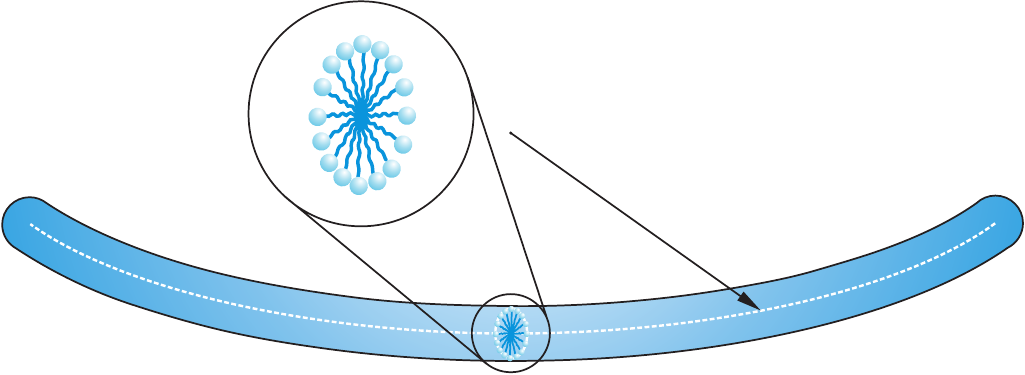}
  \put(-60,27){$R$}
  \caption{Schematic of a planar open wormlike micelle in the shape of an arc of a circle.}
   \label{f10}
\end{figure}

Consider a planar open wormlike micelle in the shape of a circular arc of length $L=R\gamma$, where $R$ and $\gamma$ respectively denote the radius of curvature and the central angle corresponding to the arc. Assume that the length of the arc is large enough, compared to the cross-sectional diameter of the wormlike micelle, to ensure assumption~\ref{as4} in Section~\ref{sectmain2}. Since the torsion $\tau$ of a circular arc vanishes, the corresponding net free-energy $E_{\text{tot}}$ in~\eqref{open WM} simplifies to
\begin{equation}
\label{openenrgy}
E_{\text{tot}}=\int_0^L(\psi_{\circ}+k_1\kappa^2)\ds+2E_{\text{cap}}.
\end{equation}
Assuming $\psi_\circ$ and $k_1$ to be uniform along ${\cal C}$, while bearing in mind that the curvature $\kappa$ is $1/R$ along the arc and $\ds=\dtheta/\kappa=R\dtheta$, \eqref{openenrgy} specializes to
\begin{equation}
\label{formfinal2}
E_{\text{tot}}=\gamma R\Big(\psi_\circ+\frac{k_1}{R^2}\Big)+2E_{\text{cap}}.
\end{equation}
\begin{figure}[t]
 \centering
 \includegraphics [height=0.9in] {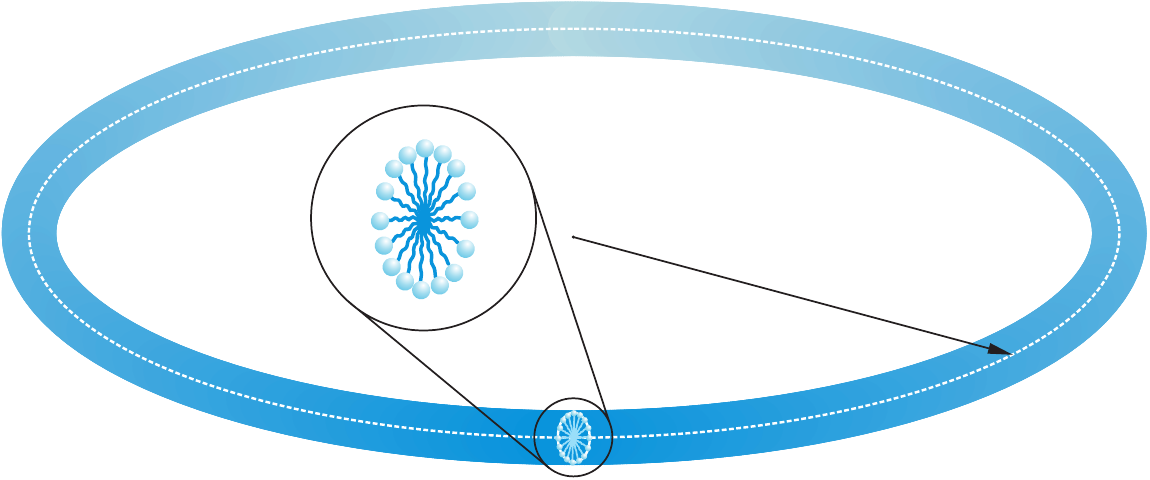}
  \put(-53,28){$R$}
 \caption{Toroidal wormlike micelle with constant radius of curvature $R$.}
  \label{f11}
\end{figure}

\subsection{Example 2: toroidal wormlike micelle}{\label{sectmain4.2}}
Consider a closed wormlike micelle in the shape of a torus with the constant major radius $R$, and the minor radius $a$, as indicated in Figure~\ref{f11}. Notice that the length of such a micelle, which is $L=2\pi R$, is assumed to be large enough relative to the minor radius $a$, so that assumption~\ref{as4} in Section~\ref{sectmain2} holds. Considering the fact that torsion $\tau$ for the centerline of a torus vanishes, and $\kappa=1/R$ along the centerline ${\cal C}$ of the torus, while keeping in mind that $\ds=R\dtheta$, the total free-energy $E_{\text{tot}}$ in \eqref{total-energy} takes the form
\begin{equation}
\label{total-energy5}
E_{\text{tot}}=\int_{0}^{2\pi}\big(\psi_{\circ}+\frac{k_1}{R^2}\big)R\dtheta.
\end{equation}
Assuming $\psi_\circ$ and $k_1$ to be uniform along ${\cal C}$, \eqref{total-energy5} results
\begin{equation}
\label{formfinal11}
E_{\text{tot}}=2\pi R\Big(\psi_\circ+\frac{k_1}{R^2}\Big).
\end{equation}

Granted that $\psi_\circ$ and $k_1$ are positive, the free energy functions expressed in~\eqref{formfinal2} and~\eqref{formfinal11} tend to $+\infty$ when $R$ approaches zero or $+\infty$. Also, the second derivative of the functions expressed in~\eqref{formfinal2} and~\eqref{formfinal11} with respect to $R$ are positive. Thus, both of the functions are convex with respect to $R$, and each possess a single minimum. As a result, the corresponding equilibrium state for each case is stable. Notice that the minimum of the total free-energy $E_{\text{tot}}$ in~\eqref{formfinal11} occurs at $R^*=\sqrt{{k_1}/{\psi_\circ}}$. Thus, our theory predicts that toroidal micelles with constant radius of curvature tend to have the major radius $R^*$ at the equilibrium state.

\subsection{A comparison between the two examples}
It has been reported that the free energy due to the end caps is relatively high (Gelbart and Ben-Shaul~\cite{gelbart1996new} and In {\etal}~\cite{in1999closed}). This relatively high free-energy might be due to the way that the surfactant molecules are packed at the ends to form two semi-spherical caps, which shield the water from the hydrocarbon chains (Gelbart and Ben-Shaul~\cite{gelbart1996new}). In equilibrium, the free energy of a collection of wormlike micelles is minimized by reducing the number of end caps (Israelachvili~\cite{israelachvili2011intermolecular}). Under certain conditions, this may be accompanied by elongation of wormlike micelles (Sharma {\etal}~\cite{sharma2009viscoelastic}). 

Consider the two examples discussed in sections~\ref{sectmain4.1} and~\ref{sectmain4.2}. As it transpires, at a specific value $\gamma_{cr}$ of $\gamma$, the total free-energy $E_{\text{tot}}$ of the open wormlike micelle in the shape of the circular arc in \eqref{formfinal2} becomes greater than the total free-energy $E_{\text{tot}}$ of the toroidal micelle in~\eqref{formfinal11}. At equilibrium, the system prefers the state with the lower free-energy. Therefore, for $\gamma>\gamma_\text{cr}$, the open wormlike micelle is expected to display a tendency to lose its end caps and form a toroidal micelle; alternatively, the micelle prefers to remain open if $\gamma$ is smaller than $\gamma_{cr}$. The angle $\gamma_{cr}$ at which this occurs, can be found by equating the right-hand sides of~\eqref{formfinal2} and~\eqref{formfinal11}:
\begin{equation}
(2\pi-\gamma_{cr})\big(\frac{k_1}{R}+\psi_{\circ}R\big)=
2E_{\text{cap}}.
\end{equation}  
Hence,
\begin{equation}
\gamma_{cr}=2\pi-\frac{2R\,E_{\text{cap}}}{k_1+\psi_{\circ}R^2}.
\end{equation}  

\section{Application of the model on toroidal micelles and comparison with previous studies}\label{sectmain6}

In this section, we expand the result of Section~\ref{sectmain4.2} by employing necessary statistical-thermodynamical concepts (Israelachvili,~\cite{israelachvili2011intermolecular,israelachvili1976theory} and Bergstr\"om~\cite{bergstrom2008thermodynamics,bergstrom1996energetics,bergstrom1996thermodynamics,bergstrom2011thermodynamics}), to make a comparison between the results of our model and the previous studies. 

\begin{figure*} [!t]
\begin{center}
\subfigure{(a)}{\resizebox*{9.3cm}{!}{\includegraphics{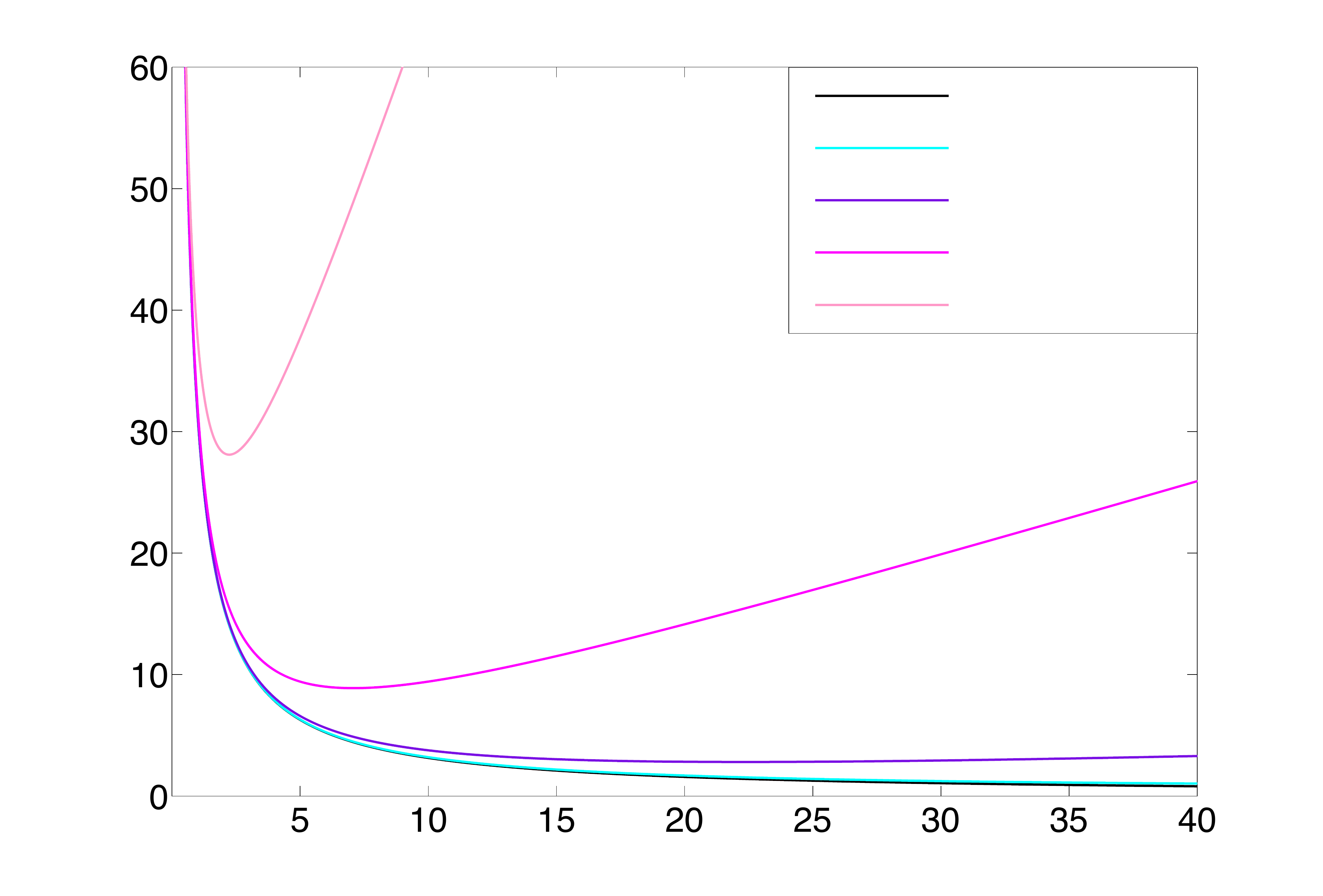}}}
 \put(-255,65){\begin{sideways}\large${E_{\text{tot}}(r)}/{k_BT}$\end{sideways}} 
\put(-148,-8){$r=R/a$}
  \put(-73,116){\small$\bar\psi_\circ=1$}
  \put(-73,126){\small$\bar\psi_\circ=0.1$}
  \put(-73,136){\small$\bar\psi_\circ=0.01$}
  \put(-73,147){\small$\bar\psi_\circ=0.001$}
  \put(-73,157){\small$\bar\psi_\circ=0.0001$}
\subfigure{(b)}{\resizebox*{9.2cm}{!}{\includegraphics{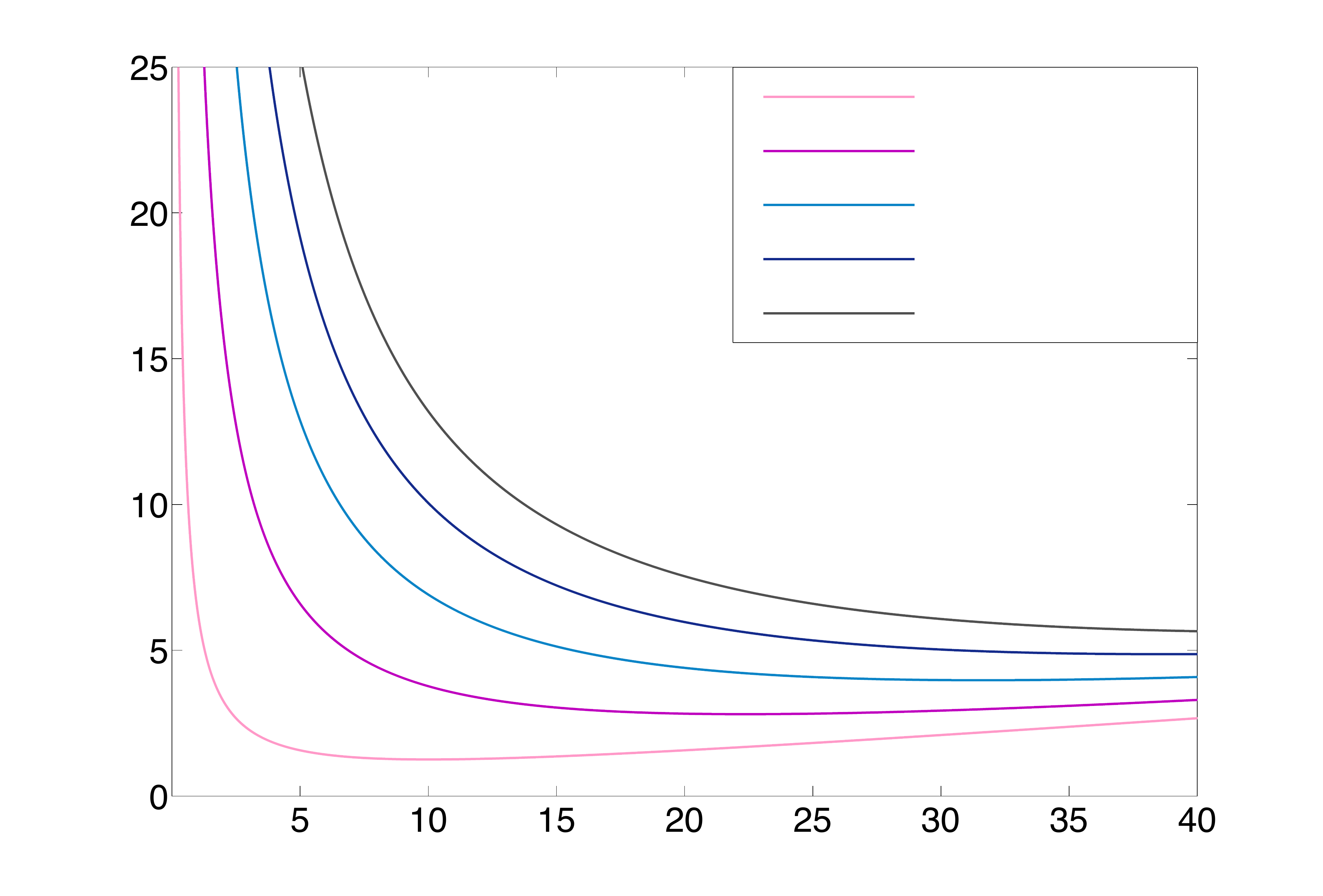}}}
   \put(-255,65){\begin{sideways}\large${E_{\text{tot}}(r)}/{k_BT}$\end{sideways}} \put(-148,-8){$r=R/a$}
  \put(-79,113){\small$k_1/a=20\,kT$}
  \put(-79,124){\small$k_1/a=15\,kT$}
  \put(-79,135){\small$k_1/a=10\,kT$}
  \put(-79,145){\small$k_1/a=5\,kT$}
  \put(-79,155){\small$k_1/a=1\,kT$}
\caption{Schematic of the dimensionless net free-energy of a toroidal micelle against the dimensionless radius $r$ of the torus, based on our theory; (a) The value of the flexural rigidity was set to $k_1/a=5\,k_BT$ (Bergstr\"{o}m~\cite{bergstrom2008thermodynamics}), and some arbitrary values for $\bar\psi_\circ=a\psi_\circ/k_BT$ were selected;
(b) The parameter $\bar\psi_\circ$ was set to $1$, and the value of $k_1/a$ was selected between 1--20 $k_BT$. The convexity of energy plots indicates the stability of the corresponding equilibrium state. Similar to the previous reports (Bergstr\"om~\cite{bergstrom2007bending}), positiveness of the flexural rigidity is a necessary condition for stability.}
\label{torus}
\end{center}
\end{figure*}
\begin{figure*}[!t]
\begin{center}
\subfigure{(a)}{\resizebox*{9.2cm}{!}{\includegraphics{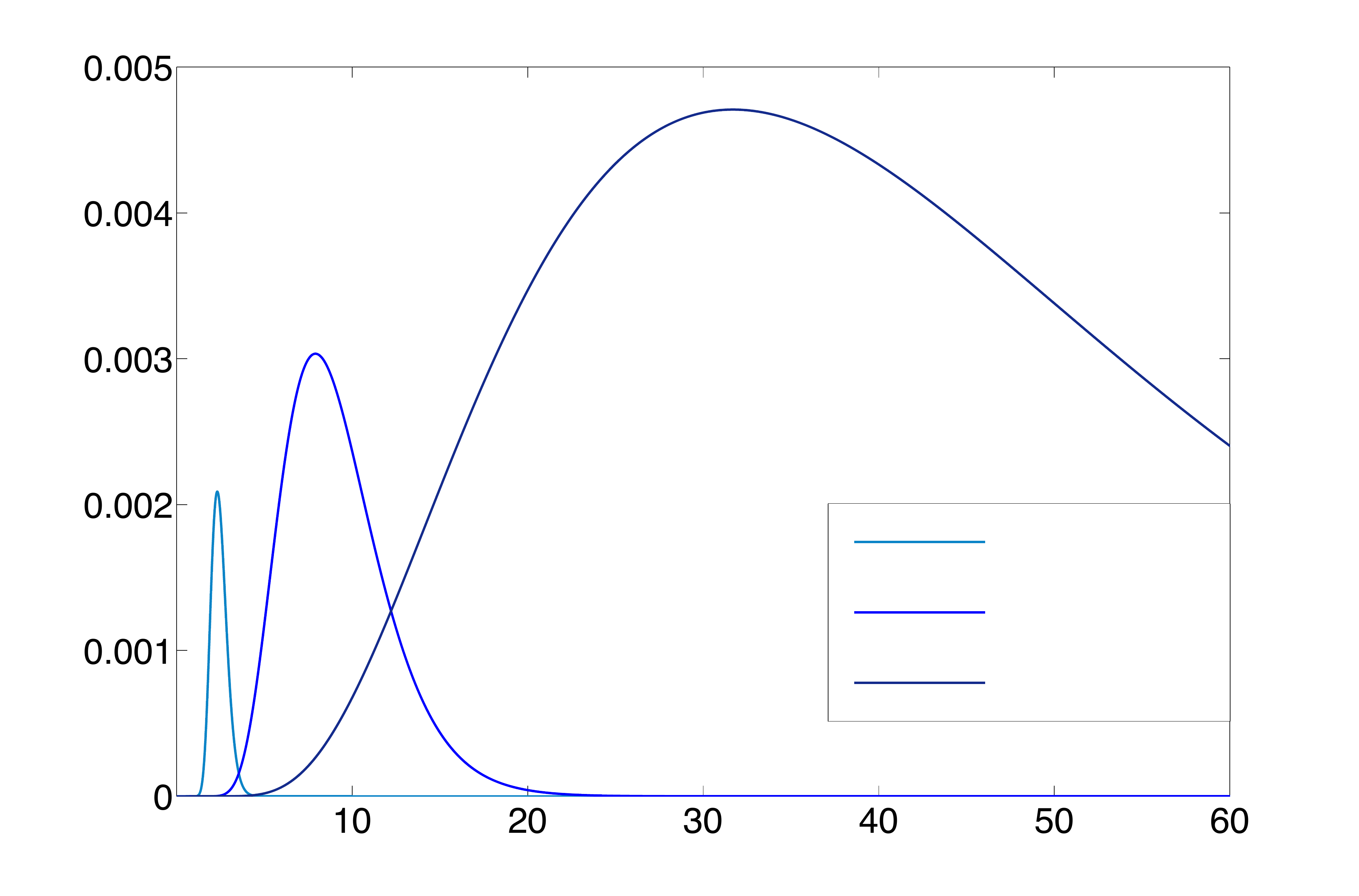}}}
 \put(-265,75){\begin{sideways}\large$\Phi_{\text{tot}}(r)$\end{sideways}} 
\put(-148,-8){$r=R/a$}
  \put(-67,40){\small$\bar\psi_\circ=0.01$}
  \put(-67,53){\small$\bar\psi_\circ=0.1$}
  \put(-67,66){\small$\bar\psi_\circ=1$}
\subfigure{(b)}{\resizebox*{9.2cm}{!}{\includegraphics{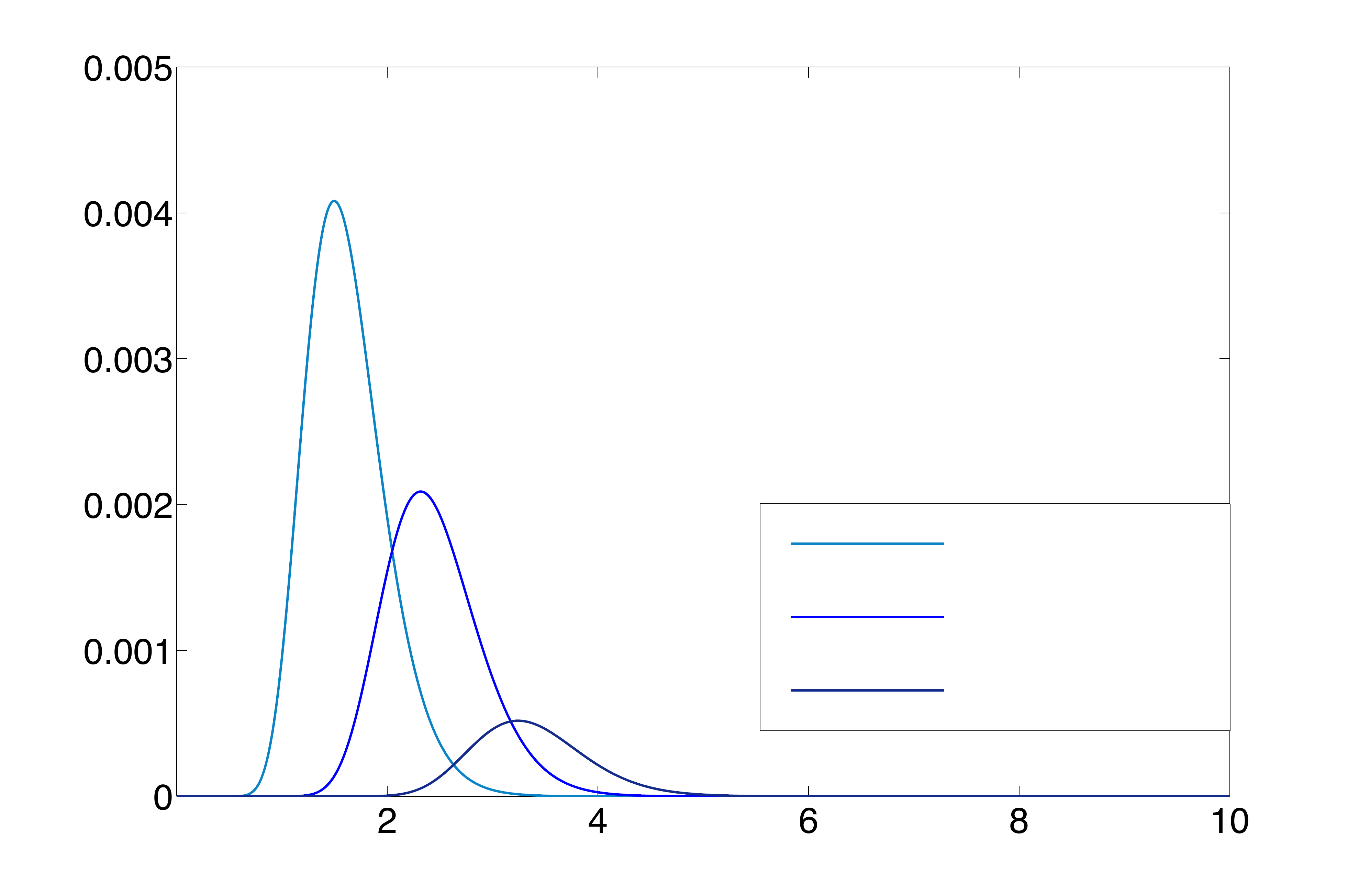}}}
   \put(-262,75){\begin{sideways}\large$\Phi_{\text{tot}}(r)$\end{sideways}}
  \put(-148,-8){$r=R/a$}
  \put(-77,37){\small$k_1/a=10kT$}
  \put(-77,52){\small$k_1/a=5kT$}
  \put(-77,66){\small$k_1/a=2kT$}
\caption{Schematic of the volume fraction density $\Phi_{\text{tot}}$ of toroidal micelles in terms of $r=R/a$, based on our model. Arbitrary factors have been used to make each set of the plots in a single frame; (a) The value of the flexural rigidity was set to $k_1/a=5\,k_BT$ (Bergstr\"{o}m~\cite{bergstrom2008thermodynamics}) and the parameter $\bar\psi_\circ=a\psi_\circ/k_BT$ was selected between 0.01--1; (b) The value of $\bar\psi_\circ$ was set to $1$, and the flexural rigidity $k_1/a$ was selected between 2--10 $k_BT$.}
\label{torussize}
\end{center}
\end{figure*}
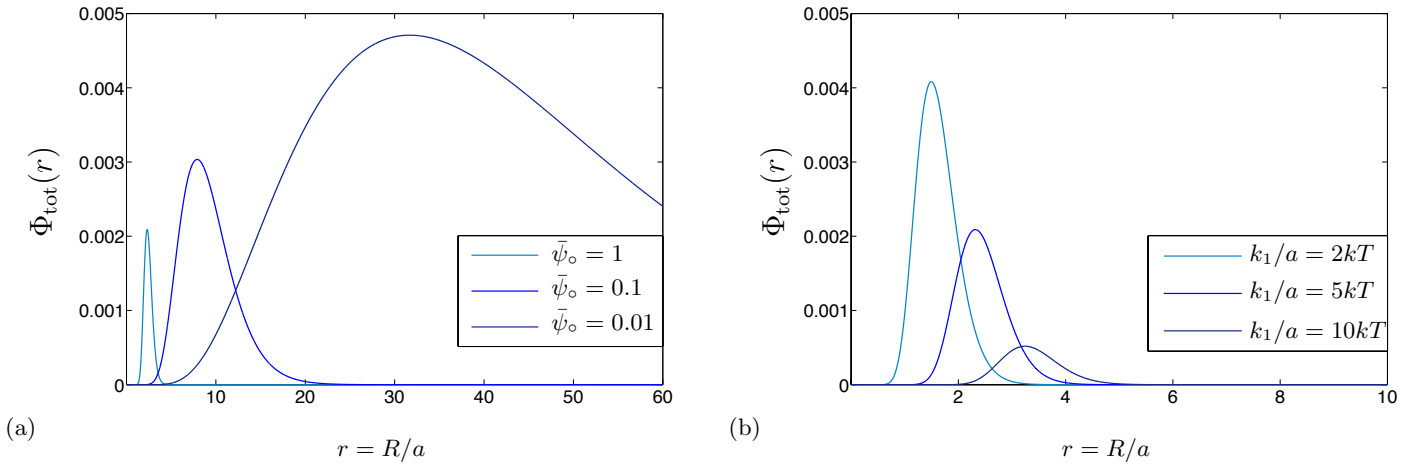

On thermodynamical grounds, the total free-energy $E_\text{tot}$ of the process of self-assembly of $N$ surfactant molecules into a single toroidal micelle can be considered as the sum of two terms: the interaction free-energy of forming a toroidal micelle out of $N$ surfactant molecules, which we denote by $N\psi_{\text{mic}}$, and the unfavourable free-energy $\Delta G_{\text{tor}}$ of self-assembling surfactants, which is the positive expression $\Delta G_{\text{tor}}=-T\Delta S_{\text{tor}}$ (Bergstr\"{o}m~\cite{bergstrom2011thermodynamics}). 
It follows from the set of thermodynamics equilibrium conditions (Bergstr\"om~\cite{bergstrom2008thermodynamics})
\begin{equation}
\label{Gibs}
\Delta G_{\text{tot}}=N\psi_{\text{mic}}+\Delta G_{\text{tor}}=0,
\end{equation}
that $\psi_{\text{mic}}$ should be a negative quantity in order for the process to be thermodynamically feasible. The ideal entropy $S_{\text{free}}$ of mixing $N$ surfactant molecules with $N_s$ solvent molecules is 
\begin{equation}
\label{entropy1}
S_{\text{free}}=-k_B\big(N\ln{\phi_{\text{free}}}+N_s\ln{\phi_s}\big),
\end{equation}
where $\phi_s$ and $\phi_{\text{free}}$ are the volume fractions of the solvent and surfactants respectively. Similarly, the entropy $S_{\text{tor}}$ of the mixture of a single toroidal micelle in $N_s$ solvent molecules is 
\begin{equation}
\label{entropy2}
S_{\text{tor}}=-k_B\big(\ln{\phi_{N}}+N_s\ln{\phi_s}\big),
\end{equation}
where $\phi_{N}$ is the volume fraction of a single toroidal micelle (Bergstr\"{o}m~\cite{bergstrom2011thermodynamics}). Considering~\eqref{entropy1} and~\eqref{entropy2}, one can simply obtain the entropy change $\Delta S_{\text{tor}}$ corresponding to the self-assembly of $N$ surfactant molecules to a single toroidal micelle as
\begin{equation}
\label{DELTAS}
\Delta S_{\text{tor}}=S_{\text{tor}}-S_{\text{free}}=-k_B\big(\ln{\phi_{N}}-N\ln{\phi_{\text{free}}}\big).
\end{equation}
In view of \eqref{DELTAS} and 
\begin{equation}
\Delta G_{\text{tor}}=-T\Delta S_{\text{tor}}=k_BT\big(\ln{\phi_{N}}-N\ln{\phi_{\text{free}}}\big),
\end{equation} 
the right-hand side equality in~\eqref{Gibs} simplifies to
\begin{equation}
\label{Gibs2}
N\psi_{\text{mic}}+k_BT\big(\ln{\phi_{N}}-N\ln{\phi_{\text{free}}}\big)=0.
\end{equation}
On introducing the net free-energy $E_{\text{tot}}$ of the formation of a toroidal micelle out of $N$ surfactant molecules as (Israelachvili~\cite{israelachvili2011intermolecular})
\begin{equation}
E_{\text{tot}}:=N\psi_{\text{mic}}-Nk_BT\ln{\phi_{\text{free}}},
\end{equation}
\eqref{Gibs2} can be expressed
alternatively as
\begin{equation}
\label{Gibs3}
E_{\text{tot}}+k_BT\ln{\phi_{N}}=0.
\end{equation}
Consequently,
\begin{equation}
\phi_N=\exp(-\frac{E_{\text{tot}}}{k_BT}).
\end{equation}
The total volume fraction $\varphi_{\text{tot}}$ may be expressed as the summation of the size distribution function (Bergstr\"om~\cite{bergstrom2011thermodynamics})
\begin{equation}
\label{volume fraction}
\varphi_{\text{tot}}=\sum_{1}^{\infty}\phi_N,
\end{equation}
which, according to Bergstr\"{o}m,~\cite{bergstrom2011thermodynamics} can be approximated by 
\begin{equation}
\label{volume fraction2}
\int_{1}^{\infty}\frac{\dN}{\dr}\exp\big(-\frac{E_{\text{tot}}(r)}{k_BT}\big)\dr.
\end{equation}
In~\eqref{volume fraction2}, $r$ denotes the ratio of the major radius $R$ of the torus, to its minor radius $a$ (see Figure~\ref{f11}). Following Bergstr\"{o}m,~\cite{bergstrom2008thermodynamics} the total volume fraction density $\Phi_{\text{tot}}$ is the integrand of the right-hand side of \eqref{volume fraction2}. Hence, the total volume fraction density (or the size distribution function) $\Phi_{\text{tot}}$ of a toroidal micelle comprised of $N$ surfactant molecules is given in terms of its net free-energy $E_{\text{tot}}(r)$ by 
\begin{equation}
\label{fractiondensity}
\Phi_{\text{tot}}=\frac{dN}{dr}\exp\Big(-\frac{E_{\text{tot}}(r)}{k_BT}\Big).
\end{equation}
The quantity $N$ can be obtained by equating the volume of the tails of $N$ surfactant molecules comprising the micelle, and that occupied by the toroidal micelle (Bergstr\"om~\cite{bergstrom2008thermodynamics}). On representing the volume of the tail of a single surfactant molecule by $v$, the volume of $N$ surfactant molecules forming the toroidal micelle is $Nv$. On the other hand, the volume of a torus with the minor radius $a$ and major radius $R$ can be obtained by applying the second Pappus--Guldinus theorem.~\cite{beyer1978crc} Hence,
\begin{equation}
\label{N}
Nv=(2\pi)(\pi a^2)R=2\pi^2a^3r,
\end{equation}
which results
\begin{equation}
\label{N2} \frac{dN}{dr}=\frac{2\pi^2a^3}{v}.
\end{equation}
Consequently,~\eqref{fractiondensity} becomes
\begin{equation}
\label{fractiondensity23}
\Phi_{\text{tot}}=\frac{2\pi^2a^3r}{v}\exp\Big(-\frac{E_{\text{tot}}(r)}{k_BT}\Big).
\end{equation}
According to our theory, the net free-energy $E_{\text{tot}}$ of a toroidal micelle is expressed in~\eqref{formfinal11}. Considering $r=R/a$,~\eqref{formfinal11} takes the form
\begin{equation}
\label{formfinal16}
E_{\text{tot}}(r)=2\pi ar\Big(\psi_\circ+\frac{k_1}{r^2a^2}\Big),
\end{equation}
or, equivalently, the dimensionless form 
\begin{equation}
\label{formfinal17}
\frac{E_{\text{tot}}(r)}{k_BT}=\frac{2\pi r}{k_BT}\Big(a\psi_\circ+\frac{k_1/a}{r^2}\Big).
\end{equation}
Figure~\ref{torus} displays the dimensionless free-energy ${E_{\text{tot}}(r)}/{k_BT}$ of a toroidal micelle against the ratio $r$. Granted that $\psi_\circ$ and $k_1$ remain positive, the net free-energy $E_\text{tot}$ is convex and possesses a minimum point corresponding to the stable equilibrium state. Further, the range of the obtained values of the minimum free-energy of a toroidal micelle in Figure~\ref{torus} is on the same order of magnitude with that obtained by Bergstr\"{o}m.~\cite{bergstrom2008thermodynamics}

Applying~\eqref{formfinal17} in~\eqref{fractiondensity23} results the final form of the volume fraction density $\Phi_{\text{tot}}$. The schematic of $\Phi_{\text{tot}}$ for a set of previously reported values of the flexural rigidity $k_1$, and some arbitrary values of the parameter $\psi_\circ$ is depicted in Figure~\ref{torussize}. As Figure~\ref{torussize}b indicates, size distribution function of toroidal micelles decreases by increasing the value of the flexural rigidity $k_1$, which is consistent with the previous findings (Bergstr\"{o}m~\cite{bergstrom2008thermodynamics}). The histogram of experimental data can be fitted to the theoretical size distribution function $\Phi_{\text{tot}}$ in~\eqref{fractiondensity23} to find the real values of the material properties $k_1$ and $\psi_\circ$. Recall that the size distribution function $\Phi_{\text{tot}}$ is a function of the total free-energy $E_{\text{tot}}$, which depends on $\psi_\circ$ and $k_1$ of the toroidal micelle. To simplify yet provide physically meaningful curve fit, by fixing the value of flexural rigidity $k_1$, one can obtain the corresponding best values of $\psi_\circ$ by fitting the size distribution function $\Phi_{\text{tot}}$ over the histogram of the experimental data. Specifically, one can rely on a range of representative values for the flexural rigidity of wormlike micelles reported in previous studies. For instance, Jung {\etal}~\cite{jung2001origins} reported the flexural rigidity in the range of 1--8 $k_BT$ for cetyltrimethylammonium bromide (CTAB)-based aggregates. More specifically, Bergstr\"om~\cite{bergstrom2008thermodynamics} set the flexural rigidity of a single toroidal micelle made of ionic surfactants to be 1--5 $k_BT$. Based on these literature reports, one may select the value of flexural rigidity for a toroidal micelle. Using such a range of representative values for the flexural rigidity, one can fit the theoretical size distribution $\Phi_{\text{tot}}$ in~\eqref{fractiondensity23} to the histogram of the experimental data to find the interval of values for $\psi_\circ$.

\section{Conclusion}\label{sectmain7}
An expression for the elastic free-energy density of wormlike micellar chains was derived taking into account interactions between the constituent surfactant molecules. The resulting expression apply only to micelles that are sufficiently long in comparison to the length of a surfactant molecule. Such micelles are modeled as a tube with circular cross-section of constant diameter whose shape is characterized by their centerlines. The free-energy density of a wormlike micelle is found to incorporate the sum of a quadratic function of the curvature and a quadratic function of the torsion of its centerline. The structure of the derived free-energy density is similar to that of free-energy density functions of polymer chains and DNA. In contrast to such models, the derived model does not exhibit intrinsic curvature nor intrinsic torsion. For a wormlike micelle consisting of more than one type of surfactant molecules, the presence of molecules with different head-group or tail conformations might lead to intrinsic curvature or intrinsic torsion. According to our modeling assumptions, such effects are ruled out.

Using the derived free-energy density, the net free-energy of either closed or open wormlike micelles were derived. Whereas the total free-energy of a closed wormlike micelle is obtained simply by integrating its free-energy density over its centerline, the total free-energy of an open wormlike micelle includes two additional contributions from its end-caps.

The derived model was applied on a special case of open wormlike micelles whose centerline has the shape of a circular arc. The model was also applied on a toruslike micelle. In both cases the minimum point corresponding to the equilibrium state was observed.

The developed model was applied to toroidal wormlike micelles, and the results were compared with those available in the literature. The range of the minimum free-energy of a toroidal micelle was found to be on the same order of magnitude as the predictions previously reported. Further, our theory predicted that the equilibrium state of toroidal micelles is stable for positive values of flexural rigidity, which is analogous to the prior predictions. Applying necessary statistical-thermodynamical concepts following previous studies, the theoretical size distribution of a toroidal micelle was found in terms of the derived free-energy. Consistent with the previous observations, it was found that the size distribution of the toroidal micelle decreases by increasing the value of its flexural rigidity. Fitting the theoretical size distribution to the experimental data, the intervals of values for material parameters can be found.

The derived model is capable of quantifying observations of wormlike micelles. Comparison of the results with previous studies on wormlike micelles available in the literature confirms the accuracy of the model.

\section{Acknowledgement}
Financial support from National Institutes of Health (NIDCD) grant DC 005788 is gratefully acknowledged. The author thanks Dr. Brian Seguin for helpful discussions, Mohsen Maleki for Figures~1--5, and Prof. Norman Wagner for helpful comments and introducing important references. 

\section{Appendix}

In this section, the details of the expansion of the right-hand side of~\eqref{e11} is carried out. On performing the change of variables $t-t_{\circ}=\delta s$ and defining $\bar\br(s)=\bx(t_{\circ})-\bx(t_{\circ}+\delta s)$, \eqref{e11} becomes
\begin{align}
\notag\label{e22}\psi&=\delta\int_{-t_{\circ}/\delta}^{(L-t_{\circ})/\delta}\mskip-8mu\int_{0}^{2\pi}\mskip-8mu\int_{0}^{2\pi}\hat\Omega\Big(\frac{\abs{\bar\br(s)}^2}{\delta^2},\bar\br(s)\cdot\bd(t_{\circ},\theta),\\
&\qquad\qquad\notag\bar\br(s)\cdot\be(t_{\circ}+\delta s,\eta),\bd(t_{\circ},\theta)\cdot\be(t_{\circ}+\delta s,\eta)\Big)\\
&\qquad\qquad f(t_{\circ})f(t_{\circ}+\delta s)\,\dtheta \deta \ds.
\end{align}
We wish to expend the right-hand side of \eqref{e22} in powers of $\delta$ neglecting terms of $o(\delta^2)$. 
The following abbreviations are used:
\begin{equation}
\left.
\begin{split}
\bn&:=\bn(t_{\circ}),\qquad \bt:=\bt(t_{\circ}),\qquad \bb:=\bb(t_{\circ}),\\[4pt]
f&:=f(t_{\circ}),\qquad f':=f'(t_{\circ}),\quad f'':=f''(t_{\circ}).
\end{split}
\right\}
\label{not1}
\end{equation}  
To achieve the expansion of the right-hand side of \eqref{e22}, the following expansions are used:
\begin{align}
\notag\bx(t_{\circ}+s\delta)&=\bx(t_{\circ})+(\frac{s^2\delta^2}{2}\kappa+\frac{s^3\delta^3}{6}\kappa')\bn+\frac{s^3\delta^3}{6}\kappa\tau\bb\\
&\notag\qquad+(s\delta-\frac{s^3\delta^3}{6}\kappa^2)\bt+o(\delta^3),\\
\notag\bn(t_{\circ}+s\delta)&=(1-\frac{s^2\delta^2}{2}(\kappa^2+\tau^2))\bn+(s\delta\tau+\frac{s^2\delta^2}{2}\tau')\bb\\
&\notag\qquad-(s\delta\kappa+\frac{s^2\delta^2}{2}\kappa')\bt+o(\delta^2),\\
\notag\bb(t_\circ+s\delta)&=-(s\delta\tau+\frac{s^2\delta^2}{2}\tau')\bn+(1-\frac{s^2\delta^2}{2}\tau^2)\bb\\
&\notag\qquad+\frac{s^2\delta^2}{2}\kappa\tau\bt+o(\delta^2),\\
\label{ex103}f(t_\circ+s\delta)&=f+s\delta f'+\frac{s^2\delta^2}{2}f''+o(\delta^2).
\end{align}
By neglecting the higher-order terms and performing the dot products, the arguments of $\hat\Omega$ in the right-hand side of equation \eqref{e22} become
\begin{equation}
\begin{split}
&\frac{\abs{\bx(t_{\circ})-\bx(t_{\circ}+s\delta)}^2}{{\delta}^2}=s^2+A_1{\delta^2}s^4+o(\delta^2),\\[4pt]
&(\bx(t_{\circ})-\bx(t_{\circ}+s\delta))\cdot\bd(t_{\circ},\theta)=A_2{\delta^2}s^2+o(\delta^2),\\[4pt]
&(\bx(t_{\circ})-\bx(t_{\circ}+s\delta))\cdot\be(t_{\circ}+s\delta,\eta)=A_3{\delta^2}s^2+o(\delta^2),\\[5pt]
&\bd(t_{\circ},\theta)\cdot\be(t_{\circ}+s\delta,\eta)=A+A_4{\delta} s+A_5{\delta^2} s^2+o(\delta^2),
\end{split}
\label{dir6}
\end{equation}  
where the coefficients $A_1$--$A_5$ in~\eqref{dir6} are expressed as 
\begin{align}
&\nonumber A=\cos(\theta-\eta),\qquad A_1=-\frac{1}{12}\kappa^2,\qquad
A_2=-\frac{1}{2}\kappa\cos\theta,\\
&\nonumber A_3=\frac{1}{2}\kappa\cos\eta,\quad A_4=\tau\sin(\theta-\eta),\\[4pt]
& A_5=\frac{1}{2}\big(\tau'\sin(\theta-\eta)-\kappa^2\cos\theta\cos\eta-\tau^2\cos(\theta-\eta)\big).
\label{dir7}
\end{align} 
The identities, $\bt'=\kappa\bn$, $\bn'=-\kappa\bt+\tau\bb$, and $\bb'=-\tau\bn$,
%
called the Frenet--Serret formulas, which express the derivatives of $\bt$, $\bn$, and $\bb$ with respect to the arclength $s$ in terms of the Frenet frame $\{\bt,\bn,\bb\}$ of ${\cal C}$ are used in~\eqref{ex103}.It should be noted that although the quantities $A_1$ through $A_5$ depend on $\theta$ and $\eta$, this dependence will not be denoted explicitly. Finally, put
\begin{align}
\label{not9091}
\nonumber\bar\Omega(s,\theta,\eta)&:=\hat\Omega(s^2,0,0,\cos(\theta-\eta)),\\
\nonumber\bar\Omega_{,i}(s,\theta,\eta)&:=\hat\Omega_{,i}(s^2,0,0,\cos(\theta-\eta)),\\
\nonumber\bar\Omega,_{ii}(s,\theta,\eta)&:=\hat\Omega,_{ii}(s^2,0,0,\cos(\theta-\eta)),\\
& \qquad\qquad\qquad i\in\{1,2,3,4\}.
\end{align}
Using the expansions \eqref{dir6} and the abbreviations \eqref{not9091}, while knowing ${\frac{-t_{\circ}}{\delta}}\ll{-\ell}$ and $\ell\ll\frac{L-t_{\circ}}{\delta}$ and using \eqref{ieradius}, the right-hand side of \eqref{e22} is, on neglecting terms proportional to $\delta^3$ and higher, 
\begin{align}
\label{expansion}\notag&\psi=\delta\int_{-\ell}^{\ell}\int_{0}^{2\pi}\mskip-10mu\int_{0}^{2\pi}\bar\Omega(s,\theta,\eta)f^2\, \dtheta \deta \ds\\
\notag&\qquad+\delta^3\int_{-\ell}^{\ell}\int_{0}^{2\pi}\mskip-11mu\int_{0}^{2\pi}\frac{1}{2}\bar\Omega(s,\theta,\eta)ff''s^2\, \dtheta \deta \ds\\
\notag&+\kappa^2\Big\{\delta^3\int_{-\ell}^{\ell}\int_{0}^{2\pi}\mskip-11mu\int_{0}^{2\pi}f^2s^2\Big[-\frac{1}{2}\cos\theta\cos\eta\,\bar\Omega_4(s,\theta,\eta)\\
\notag&\qquad-\frac{1}{12}\bar\Omega_1(s,\theta,\eta)s^2\Big]\, \dtheta \deta \ds\Big\}\\
&\notag+\tau^2\Big\{\delta^3\int_{-\ell}^{\ell}\int_{0}^{2\pi}\mskip-10mu\int_{0}^{2\pi}\frac{f^2s^2}{2}\Big[\sin^2(\theta-\eta)\bar\Omega_{44}(s,\theta,\eta)\\
&\qquad-{\cos(\theta-\eta)}\,\bar\Omega_4(s,\theta,\eta)\Big]\dtheta \deta \ds\Big\},
\end{align}
which is the free-energy density in terms of the curvature and torsion. The final form of the free-energy density for a wormlike micelle is thus of the form
\begin{equation}
\label{finalform}
\psi=\psi_{\circ}+k_1\kappa^2+k_2\tau^2,
\end{equation}
where
\begin{align}
\notag\label{expansion}\psi_{\circ}&=\delta\int_{-\ell}^{\ell}\int_{0}^{2\pi}\mskip-10mu\int_{0}^{2\pi}\bar\Omega(s,\theta,\eta)f^2\, \dtheta \deta \ds\\
&\notag\qquad+\delta^3\int_{-\ell}^{\ell}\int_{0}^{2\pi}\mskip-10mu\int_{0}^{2\pi}\frac{1}{2}\bar\Omega(s,\theta,\eta)ff''s^2\, \dtheta \deta \ds,\\
\notag k_1&=\delta^3\int_{-\ell}^{\ell}\int_{0}^{2\pi}\mskip-10mu\int_{0}^{2\pi}\frac{-f^2s^2}{12}\Big(6\,\cos\theta\cos\eta\,\bar\Omega_4(s,\theta,\eta)\\
&\notag\qquad+\bar\Omega_1(s,\theta,\eta)s^2\Big)\, \dtheta \deta \ds,\\
\notag k_2&=\delta^3\int_{-\ell}^{\ell}\int_{0}^{2\pi}\mskip-10mu\int_{0}^{2\pi}\frac{f^2s^2}{2}\Big(\sin^2(\theta-\eta)\bar\Omega_{44}(s,\theta,\eta)\\
&\qquad-{\cos(\theta-\eta)}\,\bar\Omega_4(s,\theta,\eta)\Big)\dtheta \deta \ds.
\end{align}
\footnotesize{
\bibliography{rsc} 
\bibliographystyle{rsc} 

\end{document}